\documentclass[twocolumn,preprintnumbers,superscriptaddress,nofootinbib,aps,prl,floatfix]{revtex4-2}
\pdfoutput=1
\usepackage{enumerate}
\usepackage{amsmath,amssymb}
\usepackage{graphicx}
\usepackage{slashed}
\usepackage{float}
\usepackage{comment}
\usepackage[dvipsnames]{xcolor}
\usepackage[normalem]{ulem}
\usepackage{subfigure,orcidlink}
\usepackage{multirow,array}
\usepackage[normalem]{ulem}
\usepackage{hyperref}
\hypersetup{%
	colorlinks = true,%
	linkcolor = Blue,%
	citecolor = Blue,%
	filecolor = Blue,%
	urlcolor = Blue%
}
\usepackage{tabularray}
\UseTblrLibrary{booktabs}

\begin{document}
	
\title{Can Dirac neutrinos destabilize $\mathcal{Z}_2$ domain wall network?}
    
\author{Debasish Borah \orcidlink{https://orcid.org/0000-0001-8375-282X}}
\email[Contact author: ]{dborah@iitg.ac.in}
\affiliation{Department of Physics, Indian Institute of Technology, Guwahati, Assam 781039, India}

\author{Partha Kumar Paul \orcidlink{https://orcid.org/0000-0002-9107-5635}}
\email[Contact author: ]{ph22resch11012@iith.ac.in}
\affiliation{Department of Physics, Indian Institute of Technology Hyderabad, Kandi, Telangana 502285, India}

\author{Narendra Sahu \orcidlink{https://orcid.org/0000-0002-9675-0484}}
\email[Contact author: ]{nsahu@phy.iith.ac.in}
\affiliation{Department of Physics, Indian Institute of Technology Hyderabad, Kandi, Telangana 502285, India}

\date{\today}
	
\begin{abstract}
In particle physics model building, a discrete $\mathcal{Z}_2$ symmetry is often spontaneously broken for phenomenological reasons. When this breaking occurs dynamically in the early Universe, stable domain wall networks are formed, which can eventually dominate the cosmic energy density. To avoid this problem, explicit $\mathcal{Z}_2$-breaking terms in the scalar potential are usually introduced in an ad hoc manner. In this Letter, we show that if the same $\mathcal{Z}_2$ symmetry is also responsible for generating light Dirac neutrino masses, such explicit breaking terms can instead arise radiatively from the particles involved in the Dirac mass generation. We find that the resulting bias term scales inversely with the cube of the Dirac neutrino mass, leading to a gravitational wave spectrum proportional to the sixth power of the Dirac neutrino mass. This establishes a nontrivial connection between the Dirac seesaw scale, the domain wall annihilation epoch, and the resulting stochastic gravitational wave signal. We further demonstrate that a wide range of Dirac seesaw scales can be probed by upcoming gravitational wave and cosmic microwave background experiments, while part of the parameter space simultaneously explains the observed baryon asymmetry via Dirac leptogenesis.
\end{abstract}	
	
\maketitle
\noindent
\textit{{Introduction:}} Discrete symmetries, such as a $\mathcal{Z}_2$, are a common ingredient in Dirac neutrino model building and are often assumed to be spontaneously broken by the vacuum expectation value (VEV) of a scalar field. However, spontaneous breaking of a discrete symmetry in the early Universe inevitably leads to the formation of domain walls (DWs), which, if stable, quickly dominate the energy density of the Universe and are therefore cosmologically unacceptable. A standard resolution is to introduce a small explicit breaking of the discrete symmetry through a bias term in the scalar potential \cite{Zeldovich:1974uw, Vilenkin:1981zs, Sikivie:1982qv, Gelmini:1988sf, Larsson:1996sp}. This lifts the degeneracy of the vacua, induces a pressure difference across the DWs, and triggers their collapse, releasing energy in the form of stochastic gravitational waves (GWs). While GW signals from collapsing DWs in Dirac neutrino frameworks have been widely studied, the origin of the required bias term is often assumed in an ad hoc manner \cite{Barman:2022yos, Barman:2023fad, King:2023cgv}. Therefore, in all these studies the connection between the Dirac neutrino and gravitational wave is not so profound\footnote{From a theoretical perspective, explicit breaking of discrete symmetries may arise from higher-dimensional operators suppressed by the scale of quantum gravity \cite{Rai:1992xw, Lew:1993yt} or from radiative corrections \cite{Zhang:2023nrs, Zeng:2025zjp, Borah:2025bfa}, or from a lepton parity~\cite{Ma:2025bjf}}.

In this Letter, we show that such bias terms can be generated radiatively by the same particles responsible for the Dirac neutrino mass. As a result, the formation and annihilation of domain walls are governed by the same sector that generates light Dirac neutrino masses, establishing a direct and predictive connection between the seesaw scale, the DW annihilation epoch, and the resulting GW spectrum. We refer to such defects as \textit{self-biased domain walls}. We show that both low- and intermediate-scale Dirac seesaw scenarios can lead to observable stochastic GW signals within the reach of upcoming experiments, while remaining consistent with cosmic microwave background (CMB) constraints. Although the GW-favored parameter space does not yield sizable dark radiation in the form of light Dirac neutrinos, the GW background itself contributes to the effective relativistic degrees of freedom and can be probed by future cosmic CMB observations.\\

\begin{figure}[H]
	\centering
	\includegraphics[scale=0.25]{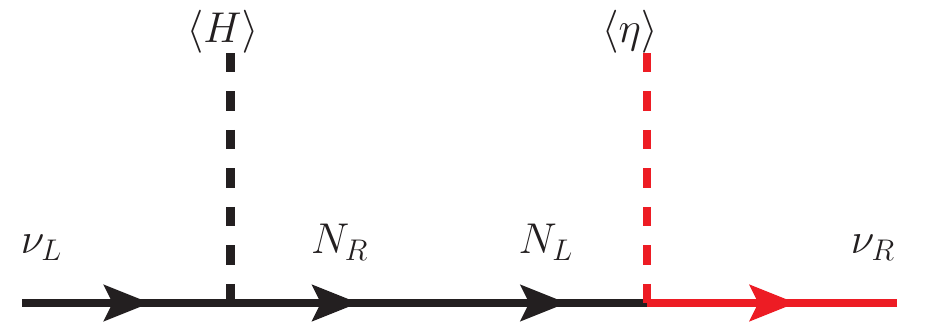}\\
	\includegraphics[scale=0.27]{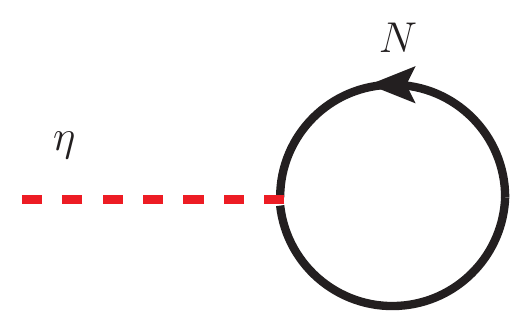}
	\includegraphics[scale=0.27]{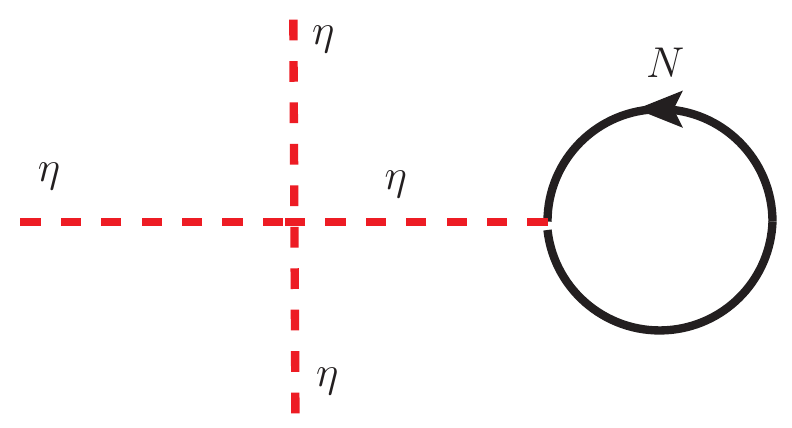}
	\caption{[\textit{Top}:] Tree-level Dirac neutrino mass. [\textit{Bottom}:] One-loop diagrams contributing to the destabilization of the domain walls.}
	\label{fig:numassbias}
\end{figure}

\noindent
\textit{{Dirac neutrino mass and domain walls:}} For the demonstration of the idea, we consider the type-I Dirac seesaw of neutrino mass where the Standard Model (SM) is extended by three generations of right-handed neutrinos $\nu_R$, a real singlet scalar $\eta$, and three copies of heavy vector-like singlet fermions $N(=N_L+N_R)$ to generate light Dirac neutrino mass. A discrete $\mathcal{Z}_2$ symmetry is imposed, under which $\nu_R$ and $\eta$ are odd while all other fields are even, thereby forbidding the Yukawa interaction $\bar{L}\tilde{H}\nu_R$. The scalar $\eta$ is allowed to acquire a non-zero vacuum expectation value $v_\eta\simeq\sqrt{\mu_\eta^2/\lambda_\eta}$, spontaneously breaking the $\mathcal{Z}_2$ symmetry. Together with electroweak symmetry breaking, this generates light Dirac neutrino masses via the diagram shown in the top panel of Fig.~\ref{fig:numassbias}. Integrating out the heavy fermions $N$, the neutrino mass is given by
\begin{eqnarray}
	m_\nu\simeq \frac{y_L y_R v_h v_\eta}{2m_N},\label{eq:numass}
\end{eqnarray}
where $v_h\simeq\sqrt{\mu_h^2/\lambda_h}$ is the Higgs vacuum expectation value. The same spontaneous breaking of the $\mathcal{Z}_2$ symmetry also leads to the formation of domain walls in the early Universe \cite{Zeldovich:1974uw, Kibble:1976sj, Vilenkin:1981zs, Saikawa:2017hiv, Roshan:2024qnv}. If stable, these domain walls rapidly dominate the energy density of the Universe, in conflict with constraints from the CMB and big bang nucleosynthesis (BBN). Assuming the walls to form after inflation, they can be eliminated by introducing a small pressure difference \cite{Zeldovich:1974uw, Vilenkin:1981zs, Sikivie:1982qv, Gelmini:1988sf, Larsson:1996sp}, commonly referred to as a bias term, which explicitly breaks $\mathcal{Z}_2$. Such bias terms are usually introduced phenomenologically as odd powers of the $\mathcal{Z}_2$-odd scalar, without any direct connection to other sectors of the theory. In this Letter, we show that the required bias term arises radiatively from Yukawa interactions involving the same heavy vector-like fermions $N$ responsible for the Dirac seesaw.\\

\noindent
\textit{{Destabilization of domain wall and gravitational waves:}} As discussed above, domain walls form when the discrete $\mathcal{Z}_2$ symmetry is spontaneously broken. These domain walls are characterized by their surface energy density, commonly referred to as the domain wall tension, which is calculated to be $\sigma_{\rm DW}=\frac{4}{3}\sqrt{\frac{\lambda_\eta}{2}}v_\eta^3\simeq\frac{2}{3}m_{\eta}v_{\eta}^2$ where $m_{\eta}=\sqrt{2\lambda_\eta}v_\eta$ with $\lambda_\eta$ being the quartic coupling in the scalar potential of $\eta$. The DW will overclose the Universe if not allowed to decay. They can be made unstable by creating a pressure difference across the two degenerate minima. This can be done by introducing arbitrary explicit $\mathcal{Z}_2$ breaking terms  in the potential. In earlier studies of DWs in Dirac seesaw models \cite{Barman:2022yos, Barman:2023fad, King:2023cgv}, such bias terms were considered independently as free parameters. Here we explore the possibility of generating all such bias terms radiatively from a single interaction of $\eta$ with the vector-like fermions $N$ which anchor the Dirac seesaw as shown in the top panel of Fig. \ref{fig:numassbias}. We introduce this $\mathcal{Z}_2$-breaking term as $y_\eta\bar{N}N\eta$. In the bottom panel of Fig. \ref{fig:numassbias}, we show the one-loop diagrams through which the $\mathcal{Z}_2$-breaking terms can be generated in the potential at the renormalizable level. We then compute the bias term which is given as
\begin{eqnarray}
    V_{\rm bias}=\frac{9m_N^3y_\eta v_\eta}{16\pi^2}\left[ 1-\gamma_E+\log\left( \frac{4\pi\Lambda^2}{m_N^2} \right) \right]\label{eq:vbias},
\end{eqnarray}
with $\gamma_E=0.577$ being the Euler constant and $\Lambda$ is the cutoff scale. It is important to note that the bias term increases with an increase in $m_N,v_\eta,y_\eta$, and it becomes negative for $m_N>4.38\Lambda$. The energy bias has to be large enough so that the DWs annihilate before they can dominate the energy density of the Universe. The wall domination criteria gives a lower bound on the bias potential to be   $ V_{\rm bias}>\frac{32\pi}{3}\mathcal{C}_{\rm ann}\frac{\mathcal{A}^2\sigma_{\rm DW}^2}{M_{\rm pl}^2}$ where $\mathcal{C}_{\rm ann}$ is a coefficient of $\mathcal{O}(1)$, $\mathcal{A}\simeq0.8\pm0.1$\cite{Hiramatsu:2013qaa} is area parameter.
The DWs must also annihilate before the BBN to be consistent with the light nuclei abundance. This puts a lower limit on the bias potential to be $V_{\rm bias}>\frac{\mathcal{C}_{\rm ann}\mathcal{A}\sigma_{\rm DW}}{\tau_{\rm BBN}}$ where $\tau_{\rm BBN}$ is the BBN time scale. It has to be noted that the bias potential $V_{\rm bias}$ can not be arbitrarily large due to the requirement of percolation of
both the vacua \cite{Gelmini:1988sf}. This gives an upper bound on the bias potential to be $V_{\rm bias}<0.795V_0$, where $V_0$ is the potential difference between the maximum and the positive ($+v_\eta$) minimum of the potential. Finally, the bias potential $V_{\rm bias}$ can be written in terms of light neutrino mass as
\begin{eqnarray}
    V_{\rm bias}=\frac{9y_L^3y_R^3y_\eta v_h^3v_\eta^4}{128\pi^2 m_\nu^3}\left[ 1-\gamma_E+\log\left( \frac{16\pi m_\nu^2\Lambda^2}{y_L^2y_R^2v_h^2v_\eta^2} \right) \right].
\end{eqnarray}

In the presence of the bias term, the domain walls become unstable and eventually collapse, releasing their energy in the form of a stochastic gravitational wave background \cite{Vilenkin:1981zs, Gelmini:1988sf, Larsson:1996sp, Hiramatsu:2013qaa, Hiramatsu:2012sc, Kadota:2015dza, Saikawa:2017hiv, Chen:2020wvu, Roshan:2024qnv,Bhattacharya:2023kws,Paul:2024iie,Ma:2025bjf}.  An additional constraint on the stochastic GW background arises from limits on new contributions to the effective number of relativistic degrees of freedom, $\Delta{N}_{\rm eff}$ or dark radiation. A sufficiently large GW energy density behaves as dark radiation and contributes to the total radiation content of the Universe, thereby modifying the expansion rate during BBN and recombination. This leads to an upper bound on the GW energy density, which can be expressed as \cite{Caprini:2018mtu}
\begin{eqnarray}
    \Omega_{\rm GW}h^2\lesssim 5.6\times10^{-6}\Delta{N}_{\rm eff}.
\end{eqnarray}

\begin{figure*}[tbh]
    \centering
    \includegraphics[scale=0.4]{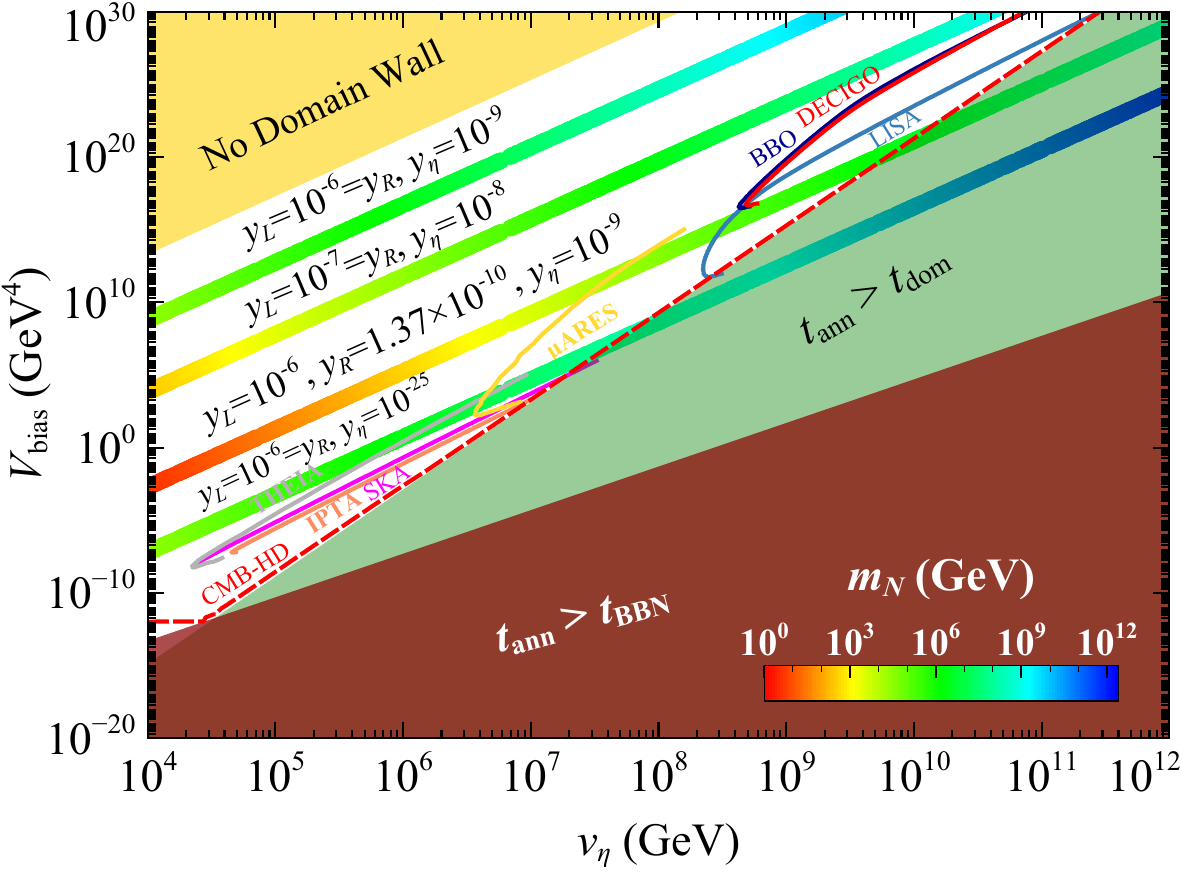}
    \includegraphics[scale=0.4]{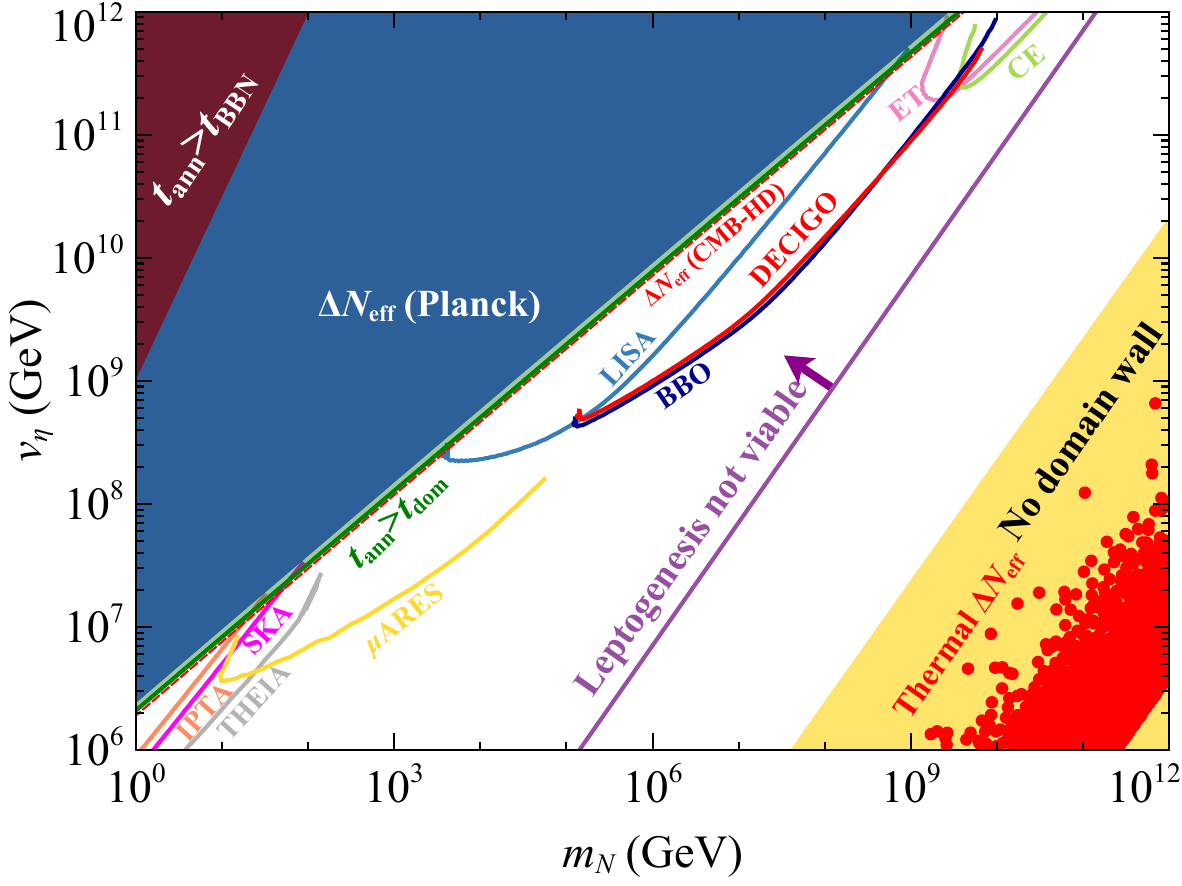}
    \caption{[\textit{Left}:] Bias potential as a function of the VEV of $\eta$ for three choices of Yukawa couplings. The color code denotes the seesaw scale or the mass of $N$. Sensitivities of different gravitational wave experiments are shown by colored contours, while the projected sensitivity of CMB-HD is indicated by a red dashed line. See main text for more details. [\textit{Right}:] VEV of $\eta$ as a function of $m_N$, obtained by fixing $\lambda_\eta = 10^{-2}$, $y_\eta = 10^{-8}$, $y_L = 10^{-7}$, $m_\nu = 0.05~{\rm eV}$, and $\Lambda = M_{\rm pl}$. $y_R$ is calculated using Eq. \eqref{eq:numass}. Sensitivity projections of different gravitational wave experiments are shown by colored contours, while exclusion limits using different criteria are indicated by shaded regions. See the main text for further details.}
    \label{fig:Vbias_plot}
\end{figure*}

In the \textit{left} panel of  Fig.~\ref{fig:Vbias_plot}, we display the region of parameter space yielding a Dirac neutrino mass $m_\nu=\mathcal{O}(0.05)~\mathrm{eV}$ in the $V_{\rm bias}$–$v_\eta$ plane. We fix $\lambda_\eta=10^{-2}$, and $\Lambda=M_{\rm pl}$. The four colored bands (top to bottom) correspond to the set of \{${y_L,y_R,y_\eta}$\} as \{$y_L=10^{-6}$, $y_R=10^{-6}$, $y_\eta=10^{-9}$\},\{$y_L=10^{-7}$, $y_R=10^{-7}$, $y_\eta=10^{-8}$\} , \{$y_L=10^{-6}$, $y_R=1.37\times10^{-10}$, $y_\eta=10^{-9}$\},  \{$y_L=10^{-6}$, $y_R=10^{-6}$, $y_\eta=10^{-25}$\}. The color coding indicates the mass $m_N$, calculated using Eq.~\eqref{eq:numass}. The dark red shaded region is excluded by BBN constraints, the green shaded region corresponds to domain walls dominating the energy density before annihilation, and the yellow shaded region denotes parameter space where domain walls do not form, which corresponds to $V_{\rm bias}>0.795V_0$. Sensitivity projections of various gravitational wave experiments BBO~\cite{Yunes:2008tw}, CE~\cite{LIGOScientific:2016wof}, ET~\cite{Punturo:2010zz}, IPTA~\cite{Hobbs:2009yy}, LISA~\cite{LISA:2017pwj}, THEIA~\cite{Garcia-Bellido:2021zgu}, DECIGO~\cite{Adelberger:2005bt}, SKA \cite{Weltman:2018zrl}, and $\mu$ARES~\cite{Sesana:2019vho} are shown by colored contours, while the projected sensitivity of CMB-HD~\cite{CMB-HD:2022bsz} is indicated by the red dashed line.

In the \textit{right} panel of  Fig.~\ref{fig:Vbias_plot}, we present the parameter space in the $v_\eta$–$m_N$ plane. We fix $\lambda_\eta=10^{-2}$, $y_\eta=10^{-8}$, $y_L=10^{-7}$, and $m_\nu=0.05~\mathrm{eV}$, while the corresponding values of $y_R$ are determined using Eq.~\ref{eq:numass}. The region where domain walls dominate the energy density before annihilation is shown by the green shaded area, whereas the dark red shaded region is excluded by BBN constraints. The yellow shaded regions denote the parameter space where domain walls do not form. The blue shaded region is excluded by Planck \cite{Planck:2018vyg} limits on $\Delta N_{\rm eff}$, while the projected sensitivity of the future CMB experiment CMB-HD is indicated by the red dashed line. Sensitivity projections of various gravitational wave experiments are shown by colored contours. In the region to the left of the magenta line, $m_N<m_\eta$, rendering leptogenesis via $N$ decay ineffective. The red points correspond to $\Delta{N}_{\rm eff}\simeq0.138$ arising from thermalization of $\nu_R$. These points, therefore, provide a complementary probe of the scenario in the absence of a detectable GW signal from DWs. 
\begin{figure}[h]
    \centering
    \includegraphics[scale=0.4]{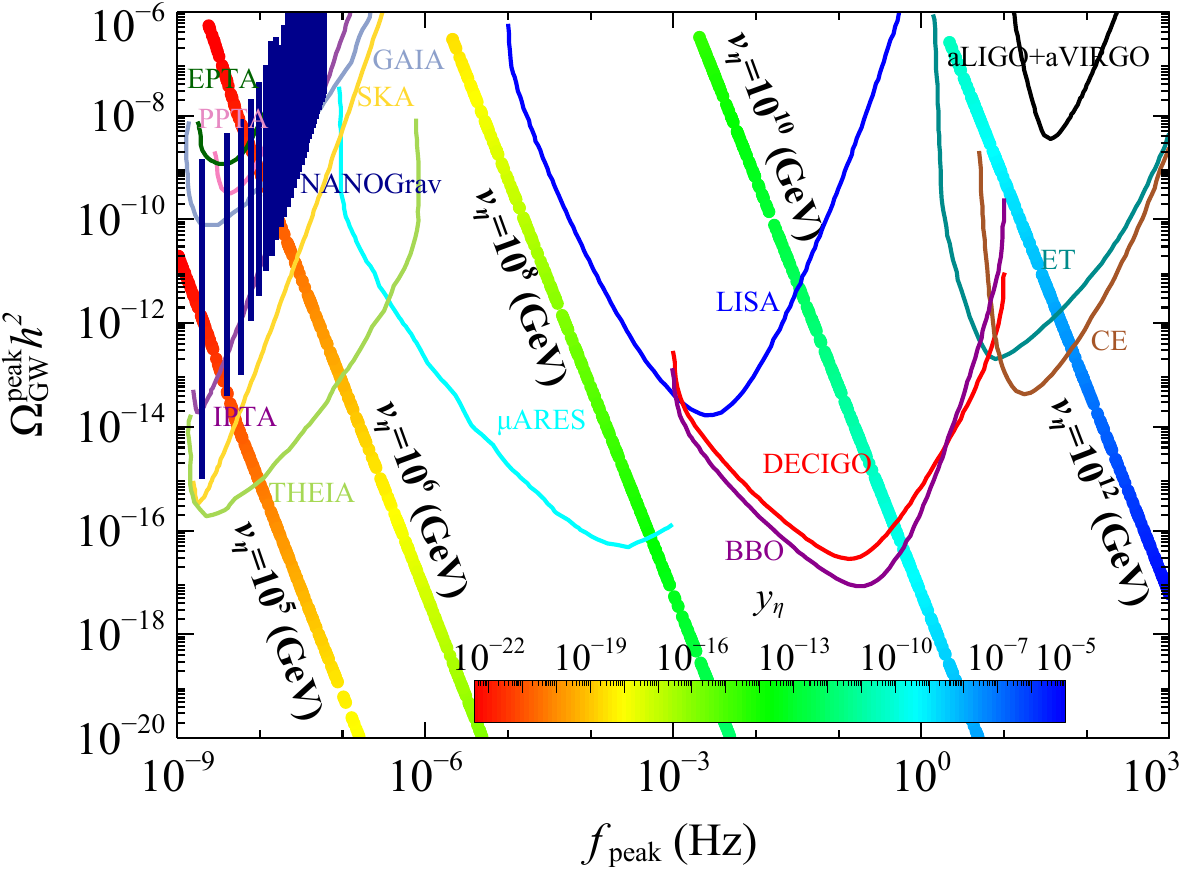}
    \caption{Parameter space in the $\Omega_{\rm GW}^{\rm peak}h^2$–$f_{\rm peak}$ plane consistent with neutrino mass is shown for five choices of $v_\eta$. $y_\eta$ is shown by the color code.  See the main text for further details.}
    \label{fig:omegaVSf}
\end{figure}

In Fig.~\ref{fig:omegaVSf}, we present the parameter space in the $\Omega_{\rm GW}^{\rm peak}h^2$–$f_{\rm peak}$ plane consistent with $m_\nu=0.05~\mathrm{eV}$. We fix $\lambda_\eta=10^{-2}$, $y_L=y_R=10^{-7}$, and $\Lambda=M_{\rm pl}$, and consider five representative values of $v_\eta$ as indicated in the figure. The Yukawa coupling $y_\eta$ is varied over the range $[10^{-5},10^{-25}]$ and is shown by the color code. The peak frequency scales as $f_{\rm peak} \propto m_N^{3/2} y_\eta v_\eta^{-1}$, while the peak amplitude behaves as $\Omega_{\rm GW}^{\rm peak} h^2 \propto v_\eta^{10} m_N^{-6} y_\eta^{-2}$. For fixed $m_N$ and $v_\eta$, decreasing $y_\eta$ shifts the peak frequency to lower values and enhances the peak amplitude, as illustrated in the figure. Such small values of $y_\eta$ are consistent with the leptogenesis parameter space, as discussed below. \\

\noindent
\textit{{Implication for leptogenesis:}} Leptogenesis can be realized in this setup through the mechanism of Dirac leptogenesis \cite{Dick:1999je, Murayama:2002je}. Since the total lepton number is conserved, CP-violating out-of-equilibrium decays of $N_1$ generate equal and opposite lepton asymmetries in the left- and right-handed sectors via the channels $N_1 \to LH$ and $N_1 \to \nu_R \eta$, respectively. Provided that the two sectors do not equilibrate, the asymmetry stored in the left-handed sector can persist and is subsequently partially converted into a baryon asymmetry by electroweak sphaleron processes. It is clear that $y_R \propto 1/v_\eta$, leading to a suppression of $y_R$ for large $v_\eta$. Since the \textit{CP} asymmetry depends on both $y_L$ and $y_R$, it becomes strongly suppressed, rendering leptogenesis unviable for $v_\eta \gtrsim 10^{4},\mathrm{GeV}$. For smaller $v_\eta$, leptogenesis is viable for $m_{N_1} \gtrsim 10^{9},\mathrm{GeV}$; however, the resulting large annihilation temperature $T_{\rm ann}$ shifts the gravitational wave peak to higher frequencies and suppresses its amplitude. This suppression can be alleviated by resonantly enhancing the \textit{CP} asymmetry through a quasi-degenerate heavy-fermion spectrum. Resonant enhancement enables successful leptogenesis for smaller $m_{N_1}$, though the VEV remains bounded by $v_\eta \lesssim 10^{5},\mathrm{GeV}$. The GW peak frequency and amplitude scale as $f_{\rm peak} \propto m_N^{3/2} y_\eta v_\eta^{-1}(\propto m_{\nu}^{-3/2}y_\eta v_{\eta}^{-1})$ and $\Omega_{\rm GW}^{\rm peak} h^2 \propto v_\eta^{10} m_N^{-6} y_\eta^{-2}(\propto v_\eta^4m_\nu^6y_\eta^{-2})$, respectively. Smaller $v_\eta$ lowers the annihilation temperature, shifting the spectrum to lower frequencies, but achieving a large low-frequency GW signal consistent with leptogenesis requires extremely small $y_\eta$. For example, choosing $y_{\eta}=10^{-25}$ and fixing the remaining parameters to \{$m_{N_{1}}=6676.6{~\rm GeV}, m_{N_{2}}-m_{N_{1}}=2.99224\times10^{-10}{~\rm GeV}, v_\eta=24805.9{~\rm GeV}, \lambda_\eta=10^{-2}, \theta_r=-0.473168+i0.625719, m_1=5.8328\times10^{-5}{~\rm eV}$\}, where $\theta_r$ denotes the rotation angle in the Casas–Ibarra parameterization of the Yukawa coupling and $m_1$ is the lightest neutrino mass, one obtains successful leptogenesis with an associated GW signal within the sensitivity ranges of THEIA and SKA.\\

\noindent
\textit{{Conclusion:}} We have proposed a novel way of domain wall disappearance in minimal Dirac neutrino mass models with additional discrete symmetries. The spontaneous breaking of $\mathcal{Z}_2$ symmetry in minimal type-I Dirac seesaw is responsible for generating light Dirac neutrino mass while also leading to the formation of domain walls. Inclusion of explicit $\mathcal{Z}_2$-breaking bias terms in the scalar potential can lead to the collapse of these walls, which also generates stochastic GWs. Instead of considering arbitrary bias terms in the potential with odd powers of the $\mathcal{Z}_2$-odd scalar, we consider the bias terms generated radiatively by the heavy vector-like singlet fermions taking part in the seesaw mechanism. We find interesting correlations among the scale of seesaw and $\mathcal{Z}_2$-breaking, light Dirac neutrino mass, as well as the GW parameters. Depending upon the scale of the seesaw, the scenario can be probed at a variety of future GW as well as CMB experiments sensitive to dark radiation in the form of gravitons or right chiral parts of Dirac neutrinos. Some part of this parameter space can also explain the observed baryon asymmetry of the Universe via Dirac leptogenesis.  \\

\noindent
\textit{{Acknowledgments.}} The authors would like to acknowledge the hospitality at IIT Hyderabad during the WHEPP 2025, where this work was initiated. P.K.P. acknowledges the Ministry of Education, Government of India, for providing financial support for his research via the Prime Minister’s Research Fellowship (PMRF) scheme.


\providecommand{\href}[2]{#2}\begingroup\raggedright\endgroup

\begin{center}
    {\bf  \large Appendix}
\end{center}
\appendix
\noindent
\textit{{Parameterization of Yukawa coupling matrices:}} The light neutrino mass matrix can be written as
\begin{eqnarray}
    m_\nu=m_L M_N^{-1}m_R,
\end{eqnarray}
where $m_L=\frac{y_Lv_h}{\sqrt{2}}$ and $m_R=\frac{y_Rv_\eta}{\sqrt{2}}$.
We can now diagonalize the matrix $m_\nu$ by \cite{Casas:2001sr,Cerdeno:2006ha} introducing two matrices $V_L$ and $V_R$ as
\begin{eqnarray}
    \hat{m}_\nu&=&V_L^\dagger m_\nu V_R,
\end{eqnarray}
where $\hat{m}_\nu={\rm diag}(m_1,m_2,m_3)$ is the diagonal light neutrino mass matrix with eigen values $m_1,m_2,m_3$.
\begin{eqnarray}
    \sqrt{\hat{m}_\nu}\sqrt{\hat{m}_\nu}&=&V_L^\dagger m_L \hat{M}_N^{-1}m_R V_R,
\end{eqnarray}
where with out loss of generality we write $M_N=\hat{M}_N={\rm diag}(m_{N_1},m_{N_2},m_{N_3})$ is the diagonal mass matrix for the fermions $N$ with eigen values $m_{N_1},m_{N_2},m_{N_3}$.
\begin{eqnarray}
    \sqrt{\hat{m}_\nu}\sqrt{\hat{m}_\nu}&=&V_L^\dagger m_L \hat{M}_N^{-1}m_R V_R
\end{eqnarray}
\begin{eqnarray}
    1&=&\left(\sqrt{\hat{m}^{-1}_\nu}V_L^\dagger m_L \sqrt{\hat{M}_N^{-1}}\right)\left(\sqrt{\hat{M}_N^{-1}}m_R V_R\sqrt{\hat{m}^{-1}_\nu}\right)\nonumber\\
    1&=&A^\dagger B
\end{eqnarray}
The left mass matrix and left sector Yukawa matrix can be written as
\begin{eqnarray}
    m_L&=&V_L\sqrt{\hat{m}_\nu}A^\dagger\sqrt{\hat{M}_N}\nonumber\\
    y_L&=&\frac{\sqrt{2}}{v_h}V_L\sqrt{\hat{m}_\nu}A^\dagger\sqrt{\hat{M}_N}\label{eq:yL}
\end{eqnarray}
Similarly, for the right-handed sector, we have
\begin{eqnarray}
    m_R&=&\sqrt{\hat{M}_N}B\sqrt{\hat{m}_\nu}V_R^\dagger\nonumber\\
    y_R&=&\frac{\sqrt{2}}{v_\eta}\sqrt{\hat{M}_N}B\sqrt{\hat{m}_\nu}V_R^\dagger\label{eq:yR}
\end{eqnarray}
For simplicity, we choose $V_L=U_{\rm PMNS}$,  $V_R=I_{3\times3}$, and $A=B=R$ and a general rotation matrix with complex rotation angle $\theta_r=a+ib$.

\noindent
\textit{Lepton asymmetry from $N$ decay:} The simultaneous decay of $N_1$ into $LH$ and $\nu_R\eta$ generates equal and opposite \textit{CP} asymmetries in the left and right-handed sectors, respectively. The asymmetries in both sectors can be partially washed out, depending on the strength of the relevant Yukawa couplings. The surviving asymmetry in the left-handed sector is subsequently converted into a baryon asymmetry through electroweak sphaleron processes. The cosmological evolution of the $N_1$ number density, along with the left and right-handed asymmetries, is described by a set of coupled Boltzmann equations (BEs), which are given by
\begin{eqnarray}
    \frac{dY_{N_1}}{dz}&=&-\frac{1}{n_\gamma \mathcal{H}z} \left( \frac{Y_{N_1}}{Y^{\rm eq}_{N_1}}-1 \right)\left[ \gamma(N_1\rightarrow LH)\right.\nonumber\\&&\left.+\gamma(N_1\rightarrow \nu_R\eta) \right]\label{eq:YN1}
\end{eqnarray}
\begin{eqnarray}
    \frac{dY_{\Delta\nu_R}}{dz}&=&\frac{1}{n_\gamma \mathcal{H}z}\left[\epsilon^1_R \left( \frac{Y_{N_1}}{Y^{\rm eq}_{N_1}}-1 \right) \gamma(N_1\rightarrow \nu_R\eta)\right.\nonumber\\&&\left.-\frac{1}{2}\frac{Y_{\Delta\nu_R}}{Y^{\rm eq}_{\nu_R}}\gamma(N_1\rightarrow \nu_R\eta)\right.\nonumber\\&&-\left. \left(\frac{Y_{\Delta\nu_R}}{Y^{\rm eq}_{\nu_R}}-\frac{Y_{\Delta L}}{Y^{\rm eq}_{L}}\right) \gamma(\nu_R\eta\rightarrow LH)  \right]\label{eq:YnuR}
\end{eqnarray}
\begin{eqnarray}
    \frac{dY_{\Delta L}}{dz}&=&\frac{1}{n_\gamma \mathcal{H}z}\left[\epsilon^1_L \left( \frac{Y_{N_1}}{Y^{\rm eq}_{N_1}}-1 \right) \gamma(N_1\rightarrow LH)\right.\nonumber\\&&\left.-\frac{1}{2}\frac{Y_{\Delta L}}{Y^{\rm eq}_{\nu_R}}\gamma(N_1\rightarrow LH)\right.\nonumber\\&&+\left. \left(\frac{Y_{\Delta\nu_R}}{Y^{\rm eq}_{\nu_R}}-\frac{Y_{\Delta L}}{Y^{\rm eq}_{L}}\right) \gamma(\nu_R\eta\rightarrow LH)  \right]\label{eq:YL}
\end{eqnarray}
where the \textit{CP} asymmetry parameters are computed to be
\begin{eqnarray}
    \epsilon^i_R\simeq\frac{1}{8\pi}\sum_{k}\frac{m_{N_i}}{m_{N_k}}\frac{{\rm Im}\left[ (y_R^\dagger y_R)_{ki} (y_L^\dagger y_L)_{ik}  \right]}{(y_R^\dagger y_R)_{ii}+(y_L^\dagger y_L)_{ii}}
\end{eqnarray}
\begin{eqnarray}
    \epsilon^i_L\simeq\frac{1}{8\pi}\sum_{k}\frac{m_{N_i}}{m_{N_k}}\frac{{\rm Im}\left[ (y_L^\dagger y_L)_{ki}(y_R^\dagger y_R)_{ik} \right]}{(y_R^\dagger y_R)_{ii}+(y_L^\dagger y_L)_{ii}}
\end{eqnarray}
\begin{figure}[h]
    \centering
    \includegraphics[scale=0.4]{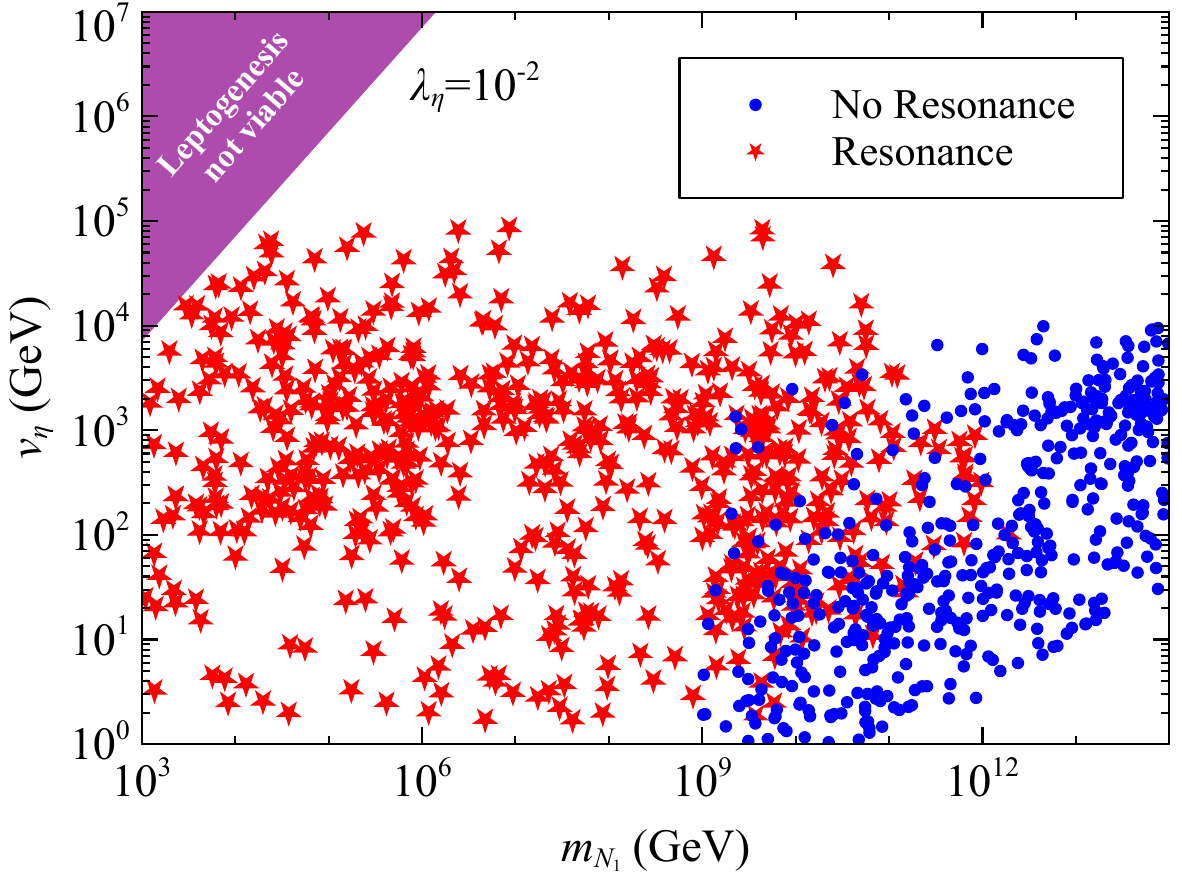}
    \caption{The parameter space giving rise to correct lepton asymmetry using the vanilla leptogenesis in the Dirac seesaw framework is shown with blue solid dots, whereas the same using the resonant leptogenesis in the Dirac seesaw framework is shown with red star points in the plane of $v_\eta$ vs $m_{N_1}$.}
    \label{fig:lepto2}
\end{figure}
In the above BEs, $\gamma(N_1 \to LH)$ denotes the reaction density for the decay of $N_1$ into the left-handed sector, while $\gamma(N_1 \to \nu_R\eta)$ corresponds to the decay into the right-handed sector. The reaction density $\gamma(\nu_R\eta \to LH)$ describes scattering processes that transfer asymmetry between the right and left-handed sectors. In Fig.~\ref{fig:lepto2}, we show the region of parameter space yielding the correct lepton asymmetry in the $v_\eta$–$m_{N_1}$ plane. The lightest neutrino mass $m_1$ and the complex rotation angle $\theta_r$ are varied randomly, while the remaining neutrino oscillation parameters are fixed to their best-fit values. The resulting viable points are shown by blue dots. From Eq.~\ref{eq:yR}, it follows that $y_R \propto 1/v_\eta$, implying that $y_R$ becomes suppressed for larger values of $v_\eta$. Although $y_L$ does not suffer from such suppression, the \textit{CP} asymmetry depends on both $y_L$ and $y_R$ and consequently becomes very small for large $v_\eta$. As a result, leptogenesis is not viable for $v_\eta \gtrsim 10^{4}~\mathrm{GeV}$.
For smaller values of $v_\eta$, leptogenesis remains viable for $m_{N_1} \gtrsim 10^{9}~\mathrm{GeV}$; however, the corresponding annihilation temperature $T_{\rm ann}$ becomes large, shifting the peak frequency of the gravitational wave spectrum to higher values. Moreover, the GW amplitude is suppressed due to the combined effects of large $T_{\rm ann}$ and small $v_\eta$. One possible way to overcome this suppression is to resonantly enhance the \textit{CP} asymmetry by considering a quasi-degenerate mass spectrum for the heavy fermions $N$. The resulting \textit{CP} asymmetries are given by
\begin{eqnarray}
\epsilon^i_R&\simeq&\sum_{k}\frac{{\rm Im}\left[ (y_R^\dagger y_R)_{ki} (y_L^\dagger y_L)_{ik}  \right]}{[(y_R^\dagger y_R)_{ii}+(y_L^\dagger y_L)_{ii}][(y_R^\dagger y_R)_{kk}+(y_L^\dagger y_L)_{kk}]}\nonumber\\&&\times\frac{(m_{N_i}^2-m_{N_k}^2)m_i\Gamma_j}{(m_{N_i}^2-m_{N_k}^2)^2+m_{N_i}^2\Gamma_j^2}
\end{eqnarray}
\begin{eqnarray}
\epsilon^i_L&\simeq&\sum_{k}\frac{{\rm Im}\left[ (y_L^\dagger y_L)_{ki}(y_R^\dagger y_R)_{ik} \right]}{[(y_R^\dagger y_R)_{ii}+(y_L^\dagger y_L)_{ii}][(y_R^\dagger y_R)_{kk}+(y_L^\dagger y_L)_{kk}]}\nonumber\\&&\times\frac{(m_{N_i}^2-m_{N_k}^2)m_i\Gamma_j}{(m_{N_i}^2-m_{N_k}^2)^2+m_{N_i}^2\Gamma_j^2}
\end{eqnarray}
In Fig.~\ref{fig:lepto2}, we show the region of parameter space yielding the correct lepton asymmetry in the $v_\eta$–$m_N$ plane for the resonant case, indicated by red stars. Resonant enhancement allows successful leptogenesis for smaller values of $m_N$; however, the VEV of $\eta$ is still restricted to $v_\eta \lesssim 10^{5}~\mathrm{GeV}$. The peak frequency of the gravitational wave spectrum scales as $f_{\rm peak} \propto m_N^{3/2} y_\eta v_\eta^{-1}$, while the peak amplitude behaves as $\Omega_{\rm GW}^{\rm peak} h^2 \propto v_\eta^{10} m_N^{-6} y_\eta^{-2}$. Consequently, smaller values of $v_\eta$ lead to a lower annihilation temperature $T_{\rm ann}$ and hence to smaller peak frequencies. Requiring consistency with successful leptogenesis while simultaneously obtaining a large GW amplitude at low frequencies necessitates extremely small values of $y_\eta$. 
\begin{figure}[h]
    \centering
    \includegraphics[scale=0.4]{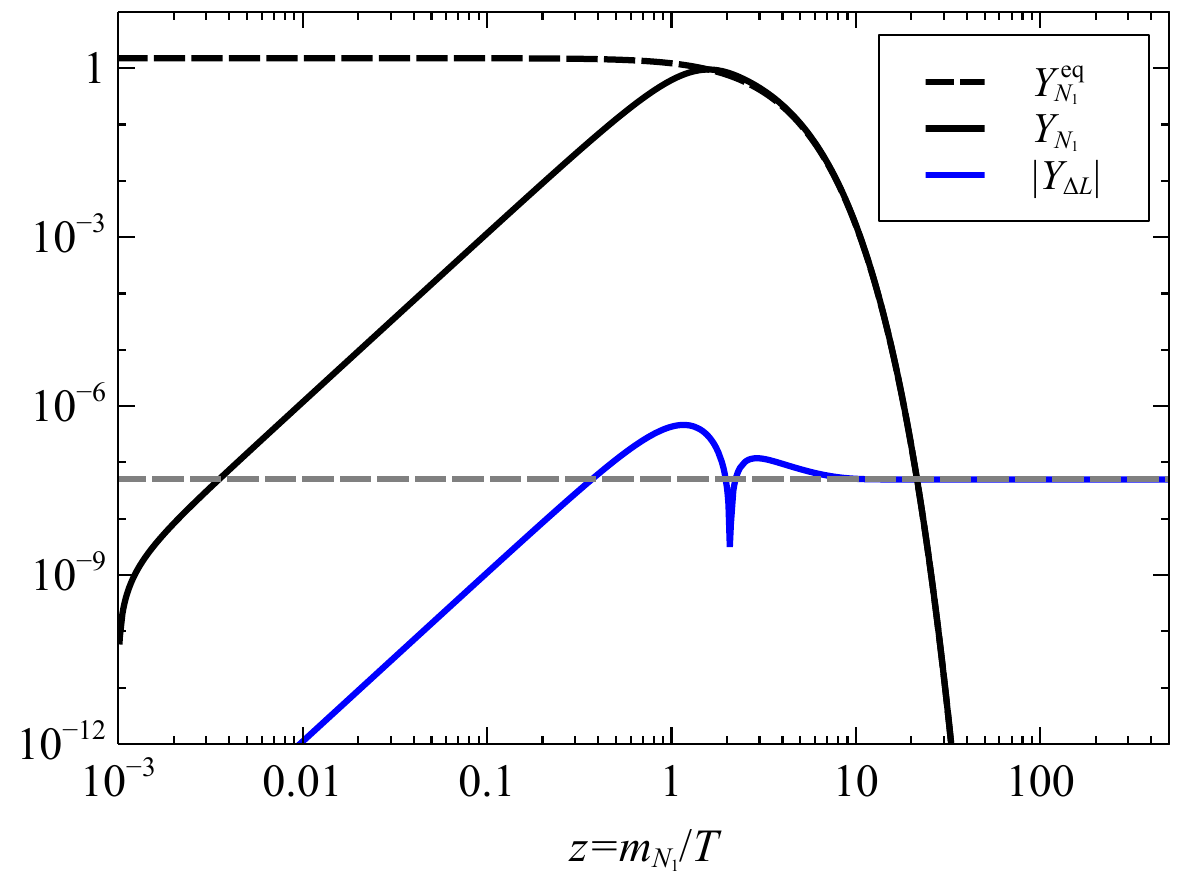}
    \caption{Evolution of comoving number densities of $N_1$ and lepton asymmetry.}
    \label{fig:lepto}
\end{figure}

Fig. \ref{fig:lepto} shows the evolution of comoving densities as a function of $z=m_{N_1}/T$ for a particular benchmark point. The blue contour indicates the evolution of left sector asymmetry, while the black solid line represents the abundance of $N_1$ and the black dashed line represents its equilibrium abundance. The horizontal gray dashed line indicates the required lepton asymmetry to produce the observed baryon asymmetry. The parameters are fixed as \{$m_{N_{1}}=6676.6{~\rm GeV},m_{N_{2}}-m_{N_{1}}=2.99224\times10^{-10}{~\rm GeV}, v_\eta=24805.9{~\rm GeV}, \lambda_\eta=10^{-2},\theta_r=-0.473168+i0.625719,m_1=5.8328\times10^{-5}$ eV\}. The \textit{CP} asymmetry parameters are calculated to be $\epsilon_L=-9.46686\times10^{-7},\epsilon_R=9.46686\times10^{-7}$. Now taking $y_{\eta}=10^{-25}$ we obtain $T_{\rm ann}=0.0506502{~\rm GeV}$. This gives an peak GW amplitude of $\Omega_{\rm GW}^{\rm peak}h^2=6.06291\times10^{-15}$ and a peak frequency $f_{\rm peak}=9.64864\times10^{-10}{~\rm Hz}$. This benchmark point lies within the sensitivity ranges of THEIA and SKA.


\end{document}